\documentclass[aps,pre,twocolumn,groupedaddress,showpacs]{revtex4}
\usepackage{graphicx}
\usepackage{amssymb}

\begin{document}
% Use the \preprint command to place your local institutional report
% number in the upper righthand corner of the title page in preprint mode.
% Multiple \preprint commands are allowed.
% Use the 'preprintnumbers' class option to override journal defaults
% to display numbers if necessary
%\preprint{}
%Title of paper
\title{Approach to jamming in an air-fluidized granular bed}
% repeat the \author .. \affiliation  etc. as needed
% \email, \thanks, \homepage, \altaffiliation all apply to the current
% author. Explanatory text should go in the []'s, actual e-mail
% address or url should go in the {}'s for \email and \homepage.
% Please use the appropriate macro foreach each type of information

% \affiliation command applies to all authors since the last
% \affiliation command. The \affiliation command should follow the
% other information
% \affiliation can be followed by \email, \homepage, \thanks as well.
\author{A.R. Abate and D.J. Durian}
%\email[]{Your e-mail address}
%\homepage[]{Your web page}
%\thanks{}
%\altaffiliation{}
\affiliation{Department of Physics \& Astronomy, University of
Pennsylvania, Philadelphia, PA 19104-6396}

%Collaboration name if desired (requires use of superscriptaddress
%option in \documentclass). \noaffiliation is required (may also be
%used with the \author command).
%\collaboration can be followed by \email, \homepage, \thanks as well.
%\collaboration{}
%\noaffiliation

\date{\today}

\begin{abstract}
Quasi-2D bidisperse amorphous systems of steel beads are fluidized
by a uniform upflow of air, so that the beads roll on a horizontal
plane.  The short-time ballistic motion of the beads is stochastic,
with non-Gaussian speed distributions and with different average
kinetic energies for the two species.  The approach to jamming is
studied as a function of increasing bead area fraction and also as a
function of decreasing air speed.  The structure of the system is
measured in terms of both the Voronoi tessellation and the pair
distribution function.  The dynamics of the system is measured in
terms of both displacement statistics and the density of vibrational
states.  These quantities all exhibit tell-tale features as the
dynamics become more constrained closer to jamming. Though the
system is driven and athermal, the behavior is remarkably
reminiscent of that in dense colloidal suspensions and supercooled
liquids.
\end{abstract}

% insert suggested PACS numbers in braces on next line
% Glass transitions, 64.70.Pf
% Fluidized beds, 47.55.Lm
% Granular systems, 45.70.–n
% 61.43.Fs Glasses
% 47.27.Sd Turbulence generated noise

% insert suggested keywords - APS authors don't need to do this
%\keywords{}

%\maketitle must follow title, authors, abstract, \pacs, and \keywords
\maketitle

% body of paper here - Use proper section commands
% References should be done using the \cite, \ref, and \label commands
%\section{}
% Put \label in argument of \section for cross-referencing
%\section{\label{}}
%\subsection{}
%\subsubsection{}

% If in two-column mode, this environment will change to single-column
% format so that long equations can be displayed. Use
% sparingly.
%\begin{widetext}
% put long equation here
%\end{widetext}

%=========================================================================================
%\section{Introduction}

One of the grand challenges in physics today is to understand
non-equilibrium systems, which evolve with time or remain in a
steady state by injection of energy.  The concept of jamming is
helping to unify such seemingly disparate non-equilibrium systems as
supercooled liquids and dense collections of droplets, bubbles,
colloidal particles, grains, and traffic~\cite{CatesPRL98,
LiuNagelN98, LiuNagelBOOK}.  In all cases, the individual units can
be jammed -- stuck essentially forever in a single packing
configuration -- by either lowering the temperature, increasing the
density, or decreasing the driving.  But what changes occur in the
structure and dynamics that signal the approach to jamming?  Which
features are generic, and which depend on details of the system or
details of the driving?  {\it A priori} all non-equilibrium systems
are different and all details should matter; therefore, the ultimate
utility and universality of the jamming concept is not at all
obvious.

Granular materials have a myriad of occurrences and applications,
and are being widely studied as idealized non-thermal systems that
can be unjammed by external forcing~\cite{Nedderman, JNB, Duran}.
Injection of energy at a boundary, either by shaking or shearing,
can induce structural rearrangements and cause the grains to explore
different packing configurations; however, the microscopic
grain-scale response is not usually homogeneous in space or time.
This can result in fascinating phenomena such as pattern formation,
compaction, shear banding, and avalanching, which have been
explained by a growing set of theoretical models with disjoint
underlying assumptions and ranges of applicability. To isolate and
identify the universally generic features of jamming behavior it
would be helpful to explore other driving mechanisms, where the
energy injection is homogeneous in space and time.  One approach is
gravity-driven flow in a vertical hopper of constant cross-section,
where flow speed is set by bottom opening. Diffusing-wave
spectroscopy and video imaging, combined~\cite{MenonSci97}, reveal
that the dynamics are ballistic at short times, diffusive at long
times, and subdiffusive at intermediate times when the grains are
`trapped' in a cage of nearest neighbors.  Such a sequence of
dynamics is familiar from thermal systems of glassy liquids and
dense colloids~\cite{LiuNagelBOOK}. Another approach is
high-frequency vertical vibration of a horizontal granular
monolayer.  For dilute grains, this is found to give Gaussian
velocity statistics, in analogy with thermal
systems~\cite{BaxterNature03}.  For dense monodisperse grains, this
is found to give melting and crystallization behavior also in
analogy with thermal systems~\cite{UrbachPRL05, Shattuck0603408}.

In this paper, we explore the universality of the jamming concept by
experimental study of disordered granular monolayers.  To achieve
homogeneous energy injection we employ a novel approach in which
grain motion is excited by a uniform upflow of air.  For dilute
grains, the shedding of turbulent wakes was found earlier to cause
stochastic motion that could be described by a Langevin equation
respecting the Fluctuation-Dissipation Theorem, in analogy with
thermal systems~\cite{RajeshNature04, RajeshPRE05, AdamPRE05}. Now
we extend this approach to dense collections of grains.  To prevent
crystallized domains, and hence to enforce homogeneous disorder, we
use a bidisperse mixture of two grain sizes.  The extent of grain
motion is gradually suppressed, and the jamming transition is thus
approached, by both raising the packing fraction and by decreasing
the grain speeds in such as way as to approach Point-J
\cite{OHernPRL02, OHernPRE03} in the jamming phase diagram. For a
given state of the system, we thoroughly characterize both structure
and dynamics using a broad set of statistical measures familiar from
study of molecular liquids and colloidal suspensions. In addition to
such usual quantities as coordination number, pair-distribution
function, mean-squared displacement, and density of states, we also
use more novel tools such as a shape factor for the Voronoi cells
and the kurtosis of the displacement distribution. We shall show
that the microscopic behavior is not in perfect analogy with thermal
systems. Rather, the two grain species have different average
kinetic energies, and their speed distributions are not Gaussian.
Nevertheless, the systematic change in behavior on approach to
jamming is found to be in good analogy with thermal systems such as
supercooled liquids and dense colloidal suspensions. Our findings
support the universality of the jamming concept, and give insight as
to which aspects of granular behavior are generic and which are due
to details of energy injection.

%=========================================================================================
\section{Methods}

The primary granular system under investigation consists of a 1:1
bidisperse mixture of chrome-coated steel ball-bearings with
diameters of $d_b=11/32~{\rm inch}=0.873$~cm and $d_s=1/4~{\rm
inch}=0.635$~cm; the diameter ratio is 1.375; the masses are 2.72~g
and 1.05~g, respectively. These beads roll on a circular horizontal
sieve, which is 6.97~inches in diameter and has a $100~\mu$m mesh
size. The packing fraction, equal to the fraction of projected area
occupied by the entire collection of beads, is varied across the
range $0.487<\phi<0.826$ by taking the total number of beads across
the range $262<N<444$.

The motion of the beads is excited by a vertical upflow of air
through the mesh at fixed superficial speed $950~\pm10$~cm/s. This
is the volume per time of air flow, divided by sieve area; the air
speed between the beads is greater according to the value of $\phi$.
The air speed is large enough to drive stochastic bead motion by
turbulence (${\rm Re}\approx10^4$), but is small enough that the
beads maintain contact with the sieve and roll without slipping. The
uniformity of the airflow is achieved by mounting the sieve atop a
$1.5\times1.5\times4~{\rm ft}^3$ windbox, and is monitored by
hot-wire anemometer.

The system of beads is illuminated by six 100~W incandescent bulbs,
arrayed in a 1~ft diameter ring located 3~ft above the sieve.
Specularly reflected light from the very top of each bead is imaged
by a digital CCD camera, Pulnix 6710, placed at the center of the
illumination ring.  The sensing element consists of a $644\times484$
array of $10\times10~\mu{\rm m}^2$ square pixels, 8~bits deep.
Images are captured at a frame rate of 120~Hz, converted to binary,
and streamed to hard-disk as AVI movies using the lossless Microsoft
RLE codec.  Run durations are 20~minutes.  The threshold level for
binary conversion is chosen so that each beads appears as a small
blob about 9 and 18 pixels in area for the small and large beads,
respectively. Note that the spot size is smaller than the bead size,
which aids in species identification and ensures that colliding
beads appear well-separated.  The minimum resolvable bead
displacement, below which there is a fixed pattern of illuminated
pixels within a blob, is about 0.1 pixels.

The AVI movies are post-processed using custom LabVIEW routines, as
follows, to deduce bead locations and speeds.  For each frame, each
bead is first identified as a contiguous blob of bright pixels. Bead
locations are then deduced from the average position of the
associated illuminated pixels.  Individual beads are then tracked
uniquely vs time, knowing that the displacement between successive
frames is always less than a bead diameter.  Finally, positions are
refined and velocities are deduced by fitting position vs time data
to a cubic polynomial.  The fitting window is $\pm5$ points, defined
by Gaussian weighting that nearly vanishes at the edges; this choice
of weighting helps ensure continuity of the derivatives.  The rms
deviation of the raw data from the polynomial fits is 0.0035~cm,
which corresponds to 0.085 pixels -- close to the minimum resolvable
bead displacement. The accuracy of the fitted position is smaller
according to the number of points in the fitting window: about
0.0035~cm/$\sqrt{5}=0.0016$~cm. This and the frame rate give an
estimate of speed accuracy as 0.2~cm/s. An alternative analysis
approach is to run the time-trace data through a low pass filter
using Fourier methods.  We find very similar results to the weighted
polynomial fits for a range of cut-off frequencies, as long as the
rms difference between raw and filtered data is between about 0.1
and 0.3 pixels.  While there is no significant difference in the
plots presented below, we slightly prefer the polynomial fit
approach based on qualitative inspection -- it is much slower but
appears better at hitting the peaks without giving spurious
oscillations.

%=======================================================================
\section{General}

\begin{figure}
\includegraphics[width=3.30in]{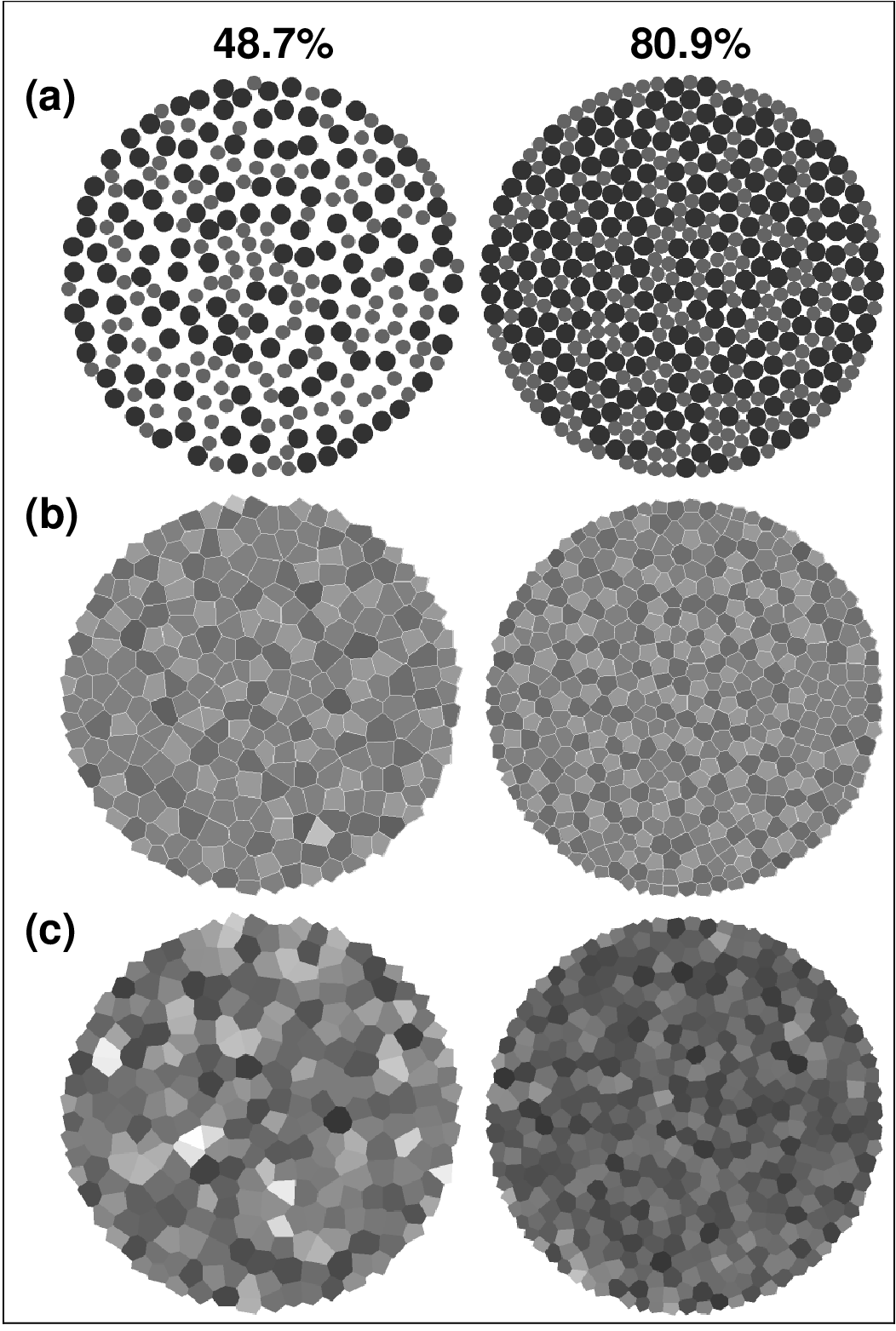}
\caption{(a) Example configurations for area fractions $\phi =
48.7\%$ and $\phi = 80.9\%$, with the big beads colored darker than
the small beads. (b-c) Voronoi tessellations for these
configurations with cells shaded darker for increasing coordination
number and circularity, respectively.}\label{VT}
\end{figure}

For orientation, Fig.~\ref{VT}a shows representative configurations
for two area fractions, $\phi=48.7\%$ and $\phi=80.9\%$.  To mimic
an actual photograph, a disk with the same diameter as the bead has
been centered over each bead's position, with a darker shade for the
big beads. Note that the configurations are disordered, and that the
two bead species are distributed evenly across the system. With
time, due to the upflow of air, the beads move about and explore
different structural configurations. Over the duration of a twenty
minute run, at our lowest packing fractions, each bead has time to
sample the entire cell several times over. The beads never
crystalize or segregate according to size.  Thus, the system appears
to be both stationary and ergodic.  However, the beads tend to idle
for a while if they come in contact with the boundary of the sieve.
Therefore, care is taken to prevent the contamination of bulk
behavior by boundary beads.

In the next sections we quantify first structure and then dynamics,
and how they both change with increasing packing fraction on
approach to jamming.  A two-dimensional random close packing of
bidisperse hard disks can occupy a range of area fractions less than
about 84\%, depending on diameter ratio and system
size~\cite{SalBook, OHernPRE03}. We find the random close packing of
our system of bidisperse beads to be at an area fraction of
$\phi_c=0.83$. If we add more beads, in attempt to exceed this
value, then there is not enough room for all beads to lie in contact
with the sieve -- some are held up into the third dimension by
enduring contact with beads in the plane.  So we expect the jamming
transition to be at or below $\phi_c=0.83$ depending on the strength
and range of bead-bead interactions. Earlier, we found that the
upflow of air creates a repulsion between two isolated beads that
can extend to many bead diameters~\cite{RajeshPRE05}. Nevertheless,
we shall show here that our system remains unjammed all the way up
to $\phi_c$ and that it develops several tell-tale signatures on
approach jamming.

%=========================================================================================
\section{Structure}

\begin{figure}
\includegraphics[width=3.30in]{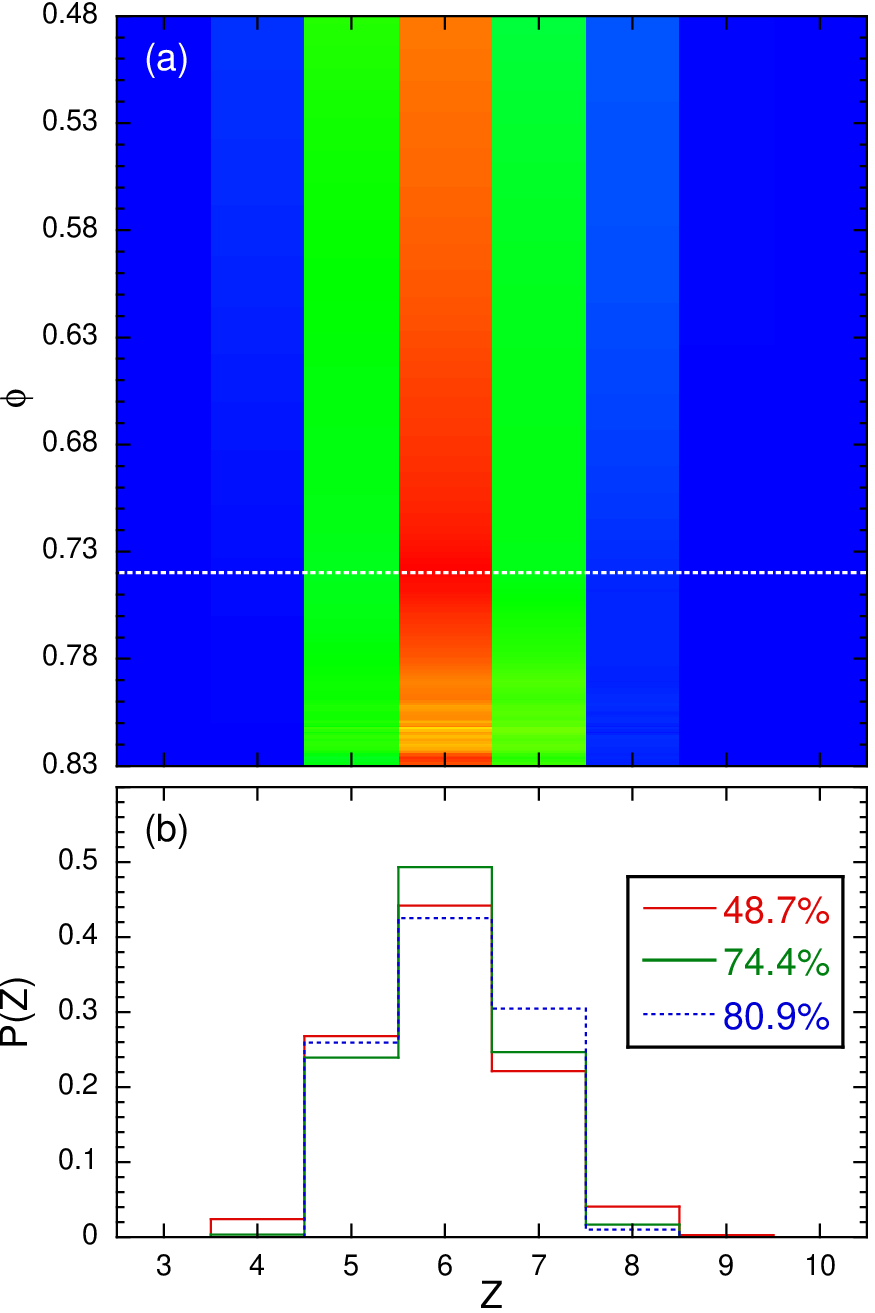}
\caption{(Color online) (a) Contour plot of the coordination number
distribution; red is for large probability density and blue is for
small.  The dashed white line indicates $\phi=0.74$. (b)
Coordination number distributions for three area fractions, as
labeled.}\label{PZ}
\end{figure}

\begin{figure}
\includegraphics[width=3.30in]{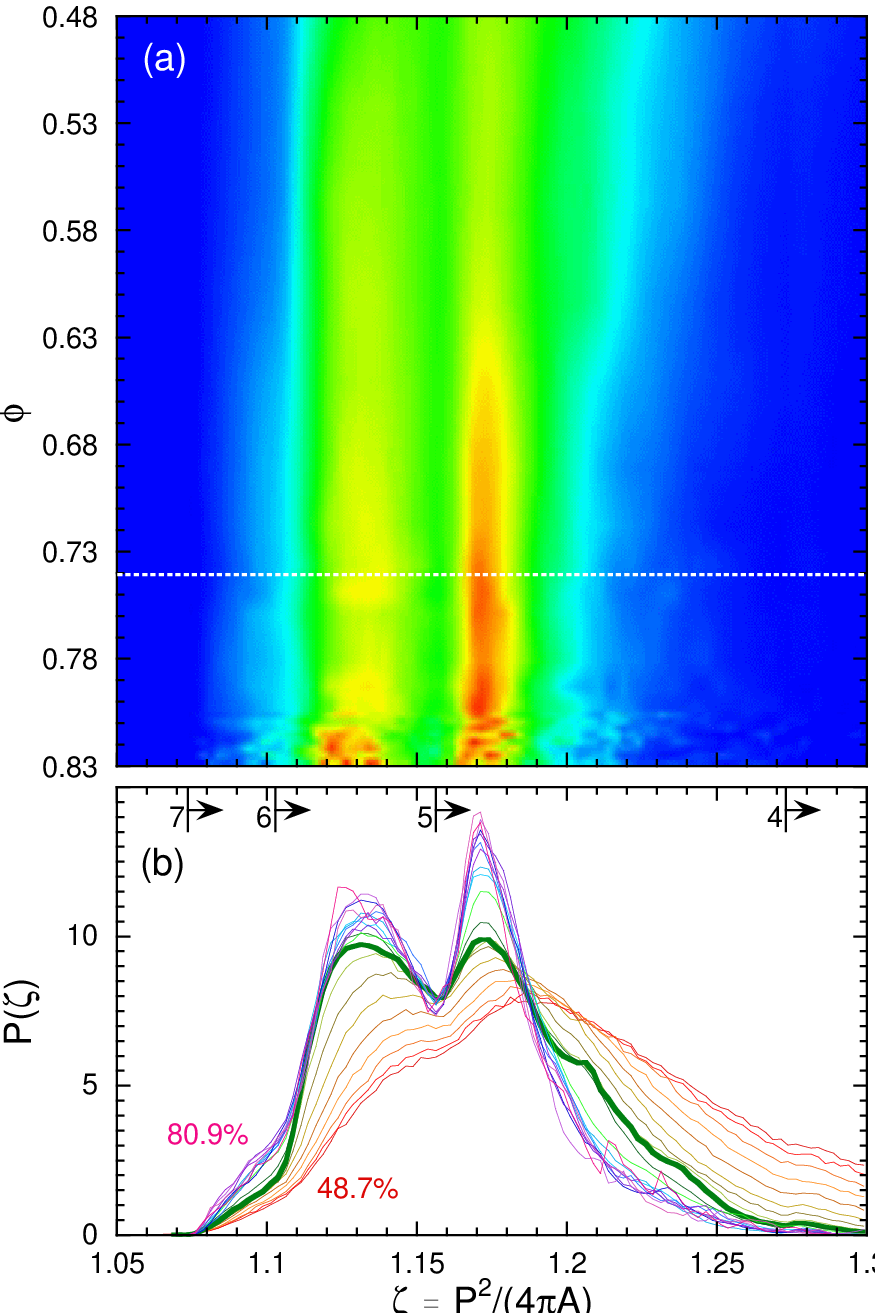}
\caption{(Color online) (a) Contour plot of the non-circularity
shape factor distributions for the Voronoi tessellation polygons.
Beyond about $74\%$ (dashed white line) a well formed second peak
develops and the distribution doesn't change much. (b) Shape factor
distributions for a sequence of area fractions; the thick green
curve is for $\phi=74.4\%$. The labels 7, 6, 5, and 4 show the
minimum shape factors for polygons with that number of
sides.}\label{PC}
\end{figure}

\begin{figure}
\includegraphics[width=3.30in]{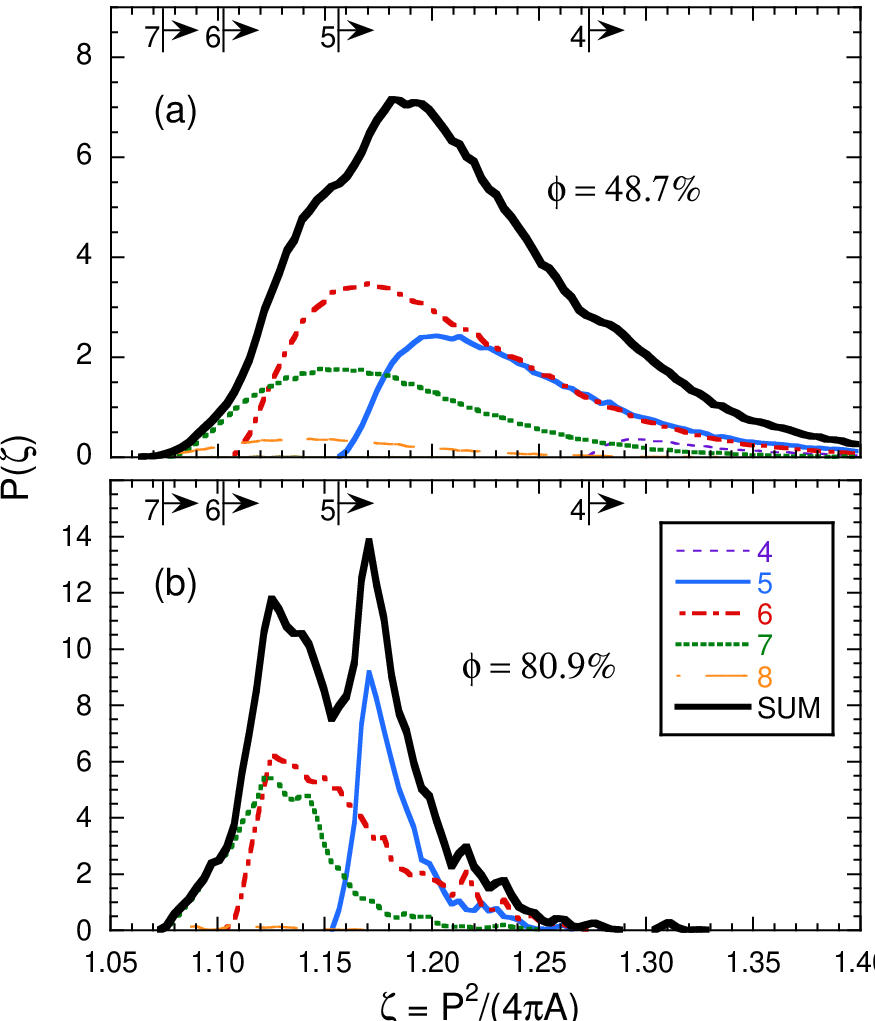}
\caption{(Color online) Non-circularity shape factor distributions
for the Voronoi tessellation polygons for (a) $\phi = 48.7\%$ and
(b) $\phi = 80.9\%$. Thick black curves are the actual
distributions, and the thin colored curves are the contributions
from cells with different numbers of sides, as labeled.}\label{PCz}
\end{figure}

\begin{figure*}
\includegraphics[width=7in]{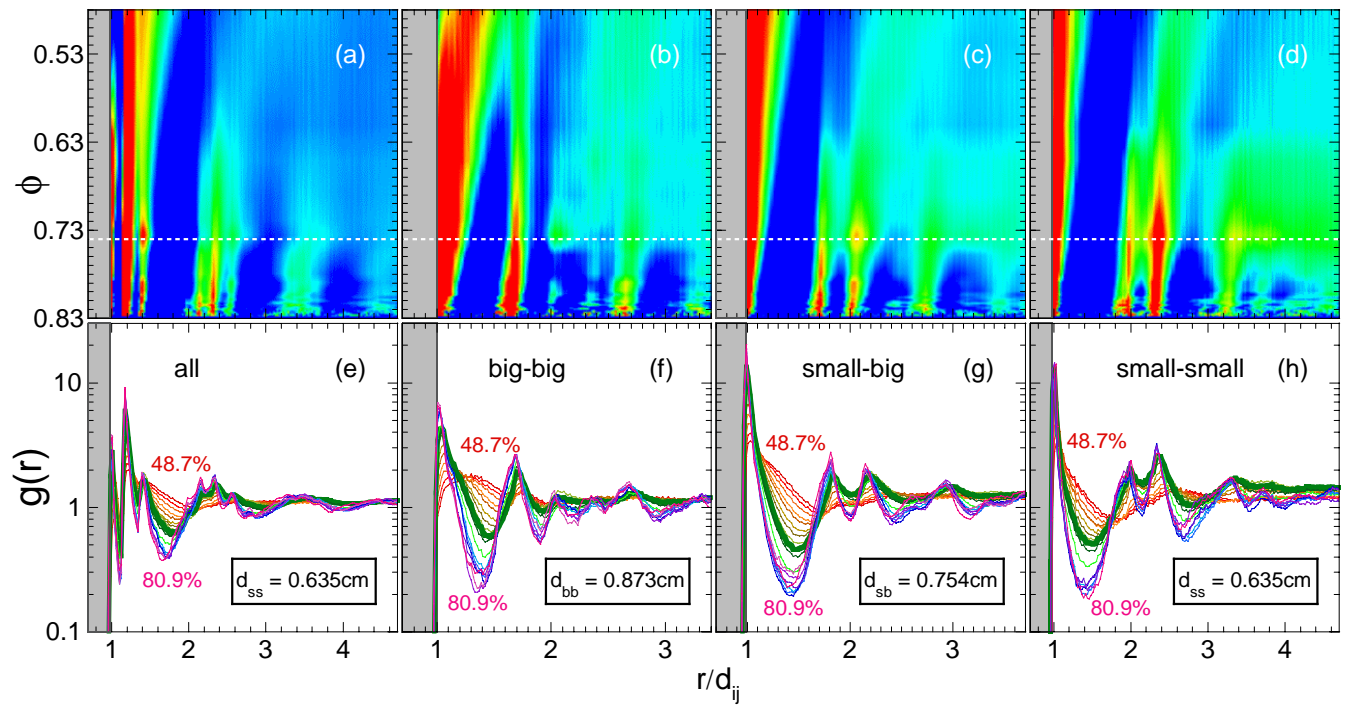}
\caption{(Color online) The radial distribution function computed
between (a,e) all beads; (b,f) big beads; (c,g) small and big beads;
and (d,h) small beads. The top row shows contour plots, where blue
is for large $g$ and red is for small; the dashed white line
represents $\phi=0.74$. The bottom row shows data curves for
different area fractions; the thick green curves are for
$\phi=0.744$. The grayed region represents the distance excluded by
hard-core contact, for $r$ less than the sum $d_{ij}$ of
radii.}\label{GrC}
\end{figure*}

\begin{figure}
\includegraphics[width=3.30in]{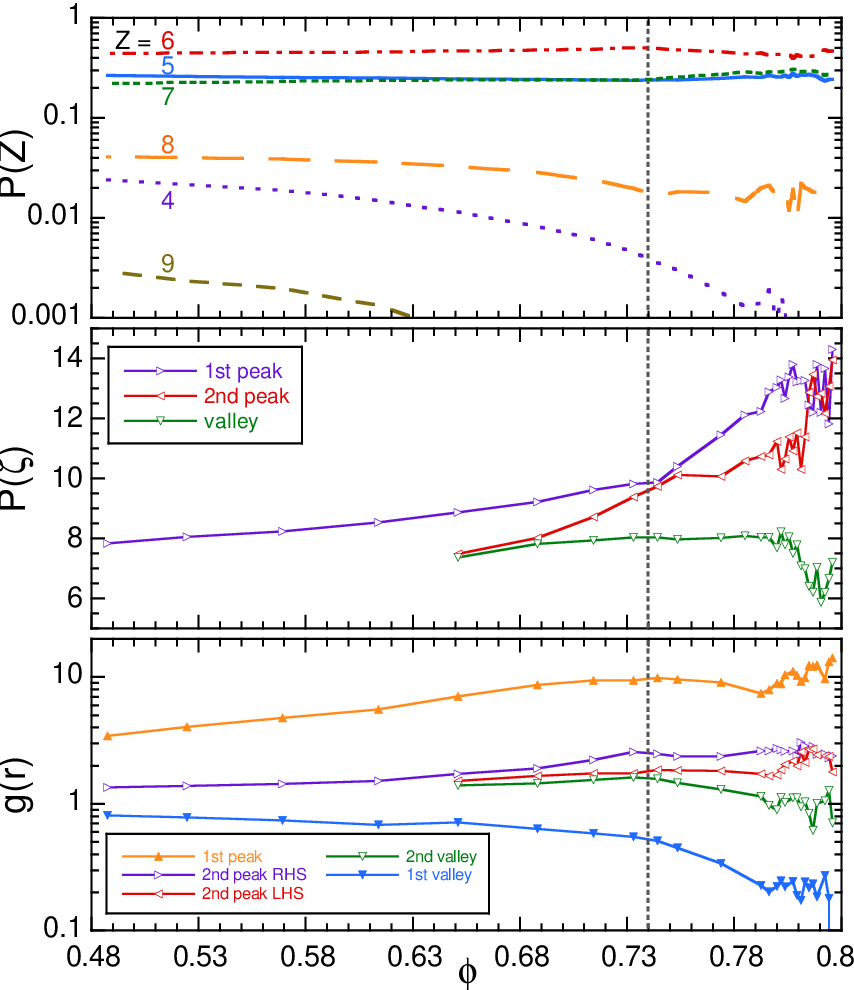}
\caption{(Color online) The area fraction dependence of (a) the
occurrence probability $P(Z)$ of Voronoi cells with $Z$ sides, and
peak and valley values of (b) the shape factor distribution and (c)
the pair distribution function for small beads.}\label{Strs}
\end{figure}

\subsection{Coordination number}

Perhaps the simplest structural quantity is the coordination number
$Z$, equal to the number of nearest neighbors for each bead.  This
can be most conveniently measured by constructing a Voronoi
tessellation, which is dual to the position representation, and by
counting the number of sides of each polygonal Voronoi cell.
Examples are shown in Fig.~\ref{VT}b for the same configurations
shown in Fig.~\ref{VT}a. Here the Voronoi cells are shaded darker
for greater numbers of sides.  The coordination number ranges
between 3 and 9, but by far the most common numbers are 5, 6, and 7
irrespective of area fraction. It seems that 5- and 7-sided cells
appear together, and that 6-sided cells sometimes appear in small
compact clusters.

The distribution $P(Z)$ of coordination numbers, and trends vs area
fraction, are displayed in Fig.~\ref{PZ}.  The results are obtained
by averaging over all times and over all beads away from the
boundary.  The bottom plot shows actual distributions for three area
fractions, two of which are the same as in Fig.~\ref{VT}. The main
effect of increasing the area fraction is to increase the fraction
of 7-sided cells at the expense of all others, until essentially
only 5-6-7 sided cells remain. This trend is more clearly displayed
in a contour plot, Fig.~\ref{PZ}a, where the value of $P(Z)$ is
indicated by color as a function of both $Z$ and area fraction
$\phi$.  For area fractions above about 74\%, indicated by a dashed
white line, the 4-, 8-, and 9-sided cells have essentially
disappeared. This effect is rather subtle, owing to the discrete
nature of the coordination number.  Six-sided cells are always the
most plentiful; their abundance gently peaks near 74\%.

\subsection{Circularity}

A more dramatic measure of structural change upon approach to
jamming is found by considering the {\it shapes} of the Voronoi
cells.  One choice for a dimensionless measure of deviation from
circularity is
\begin{equation}\label{Circ}
    \zeta = P^2/(4\pi A),
\end{equation}
where $P$ is the cell perimeter and $A$ is the cell area. This
quantity was recently used to study crystallization of
two-dimensional systems, both in simulation~\cite{MouckaPRL05} and
experiment~\cite{Shattuck0603408}.  By construction $\zeta$ equals
one for a perfect circle, and is higher for more rough or oblong
shapes; for a regular $Z$-sided polygon it is
\begin{equation}\label{CircZ}
    \zeta_Z = (Z/\pi)\tan(\pi/Z),
\end{equation}
which sets a lower bound for other $Z$-sided polygons.  As an
example in Fig.~\ref{VT}c the Voronoi cells are shaded darker for
more circular shapes, i.e.\ for those with smaller non-circularity
shape factors. Since $\zeta_Z$ decreases with increasing $Z$, it may
be expected that the shape factor is related to coordination number.
The advantage is that $\zeta$ is a continuous variable while $Z$ is
discrete.

Shape factor distributions, $P(\zeta)$, and the way they change with
increasing area fraction, are displayed in Fig.~\ref{PC}b.  These
are obtained by constructing Voronoi tessellations, and averaging
over all times and over all beads.  At low area fractions,
$P(\zeta)$ exhibits a single broad peak.  At higher area fractions,
this peaks moves to lower $\zeta$, i.e.\ to more circular domains,
and eventually {\it bifurcates} into two sharper peaks. This trend
can be seen, too, in the contour plot of Fig.~\ref{PC}a where color
indicates the value of $P(\zeta)$ as a function of both
non-circularity $\zeta$ and area fraction $\phi$. This plot shows
that the double peak becomes essentially completely developed around
$\phi=0.74$.  This is the same area fraction singled out by a subtle
change in the coordination number distribution.  Thus the shape
factor and its distribution are useful for tracking the change in
structure as a liquid-like system approaches a disordered jammed
state.

The origin of the double peak in the shape factor distribution
$P(\zeta)$ can be understood by considering the contributions
$P_Z(\zeta)$ made by cells with different coordination numbers.
These contributions are defined such that
$P(\zeta)=\sum_{Z=3}^{\infty}P_Z(\zeta)$ and
$P(Z)=\int_{\zeta_Z}^\infty P_Z(\zeta){\rm d}\zeta$; in particular,
cells with coordination number $Z$ contribute a sub-distribution
$P_Z(\zeta)$ which must vanish for $\zeta<\zeta_Z$ according to
Eq.~(\ref{CircZ}) and which subtends an area equal to the
coordination number probability. As an example, the shape factor
distribution is shown along with the individual contributions in
Fig.~\ref{PCz}. At low area fractions, in the top plot, the broad
peak in $P(\zeta)$ is seen to be composed primarily of overlapping
broad contributions from 5-, 6-, and 7-sided cells. At high area
fractions, in the bottom plot, the double peak in $P(\zeta)$ is seen
to be caused by 5-sided cells for the right peak and by overlapping
contributions of 6- and 7-sided cells for the left peak.  At these
higher area fractions, the individual contributions $P_Z(\zeta)$ are
more narrow and rise more sharply from zero for $\zeta>\zeta_Z$.  In
other words, the Voronoi cells all become more circular at higher
packing fractions. Due to disorder, there is a limit to the degree
of circularity that cannot be exceeded and so the changes in the
circularity distribution eventually saturate. For our system this
happens around $\phi=0.74$, which is well below random close packing
at $\phi_c=0.83$.

\subsection{Radial distribution}

We now present one last measure of structure that is commonly used
in amorphous systems: the radial or pair distribution function
$g(r)$. This quantity relates to the probability of finding another
bead at distance $r$ away from a given bead.  For large $r$, it is
normalized to approach one -- indicating that the system is
homogeneous at long length scales.  For small $r$ in hard-sphere
systems like ours, it vanishes for $r$ less than the sum of bead
radii. Since we have two species of beads, big and small, there are
four different distributions to consider: between any two beads,
between only big beads, between big and small beads, and between
only small beads.

Data for all four radial distribution functions are collected in
Fig.~\ref{GrC}.  As usual, these data were obtained by averaging
over all times and over all beads away from the boundary.  The
bottom row shows functions plotted vs $r$ for different area
fractions, while the top row shows contour plots where color
indicates the value of $g(r)$ as a function both $r$ and area
fraction; radial distance is scaled by the sum of bead radii
$d_{ij}$. All four radial distribution functions display a global
peak at hard core contact, $r/d_{ij}=1$.  For increasing area
fractions, these peaks become higher and more narrow, while
oscillations develop that extend to larger $r$.  Also, near
$r/d_{ij}=2$ there develops a second peak that grows and then
bifurcates into two separate peaks.  Both the growth of the peak at
$r/d_{ij}=1$, relative to the deepest minimum, and the splitting of
the peak near $r/d_{ij}=2$ have been taken as structural signatures
of the glass transition~\cite{CargillJAP70b, AbrahamPRL78,
Silbert0601012}. For our system the split second peak becomes fully
developed for area fractions greater than about $\phi=0.74$, the
same area fraction noted above with regards to changes in the
Voronoi tessellations.

\subsection{Summary}

Well-defined features develop in several structural quantities as
the area fraction increases.  The most subtle is an increase in the
number of 7-sided Voronoi cells at the expense of all other
coordination numbers, and the disappearance of nearly all 4- and
9-sized cells. The most obvious are the splitting of peaks in the
shape factor distribution and the radial distribution functions.
These key quantities are extracted from our full data sets and are
displayed vs area fraction $\phi$ in Fig.~\ref{Strs}.  The top plot
shows the fraction $P(Z)$ of Voronoi cells with $Z$ sides, for
several coordination numbers; the middle plot shows peak and valley
values of the probability density $P(\zeta)$ for Voronoi cells with
shape factor $\zeta$; the bottom plot shows peak and valley values
of the pair distribution function $g(r)$ for small beads.  These
three plots give a consistent picture that $\phi=0.74$ is the
characteristic area fraction for structural change.

%=========================================================================================
\section{Dynamics}

\subsection{Data}

\begin{figure*}
\includegraphics[width=7in]{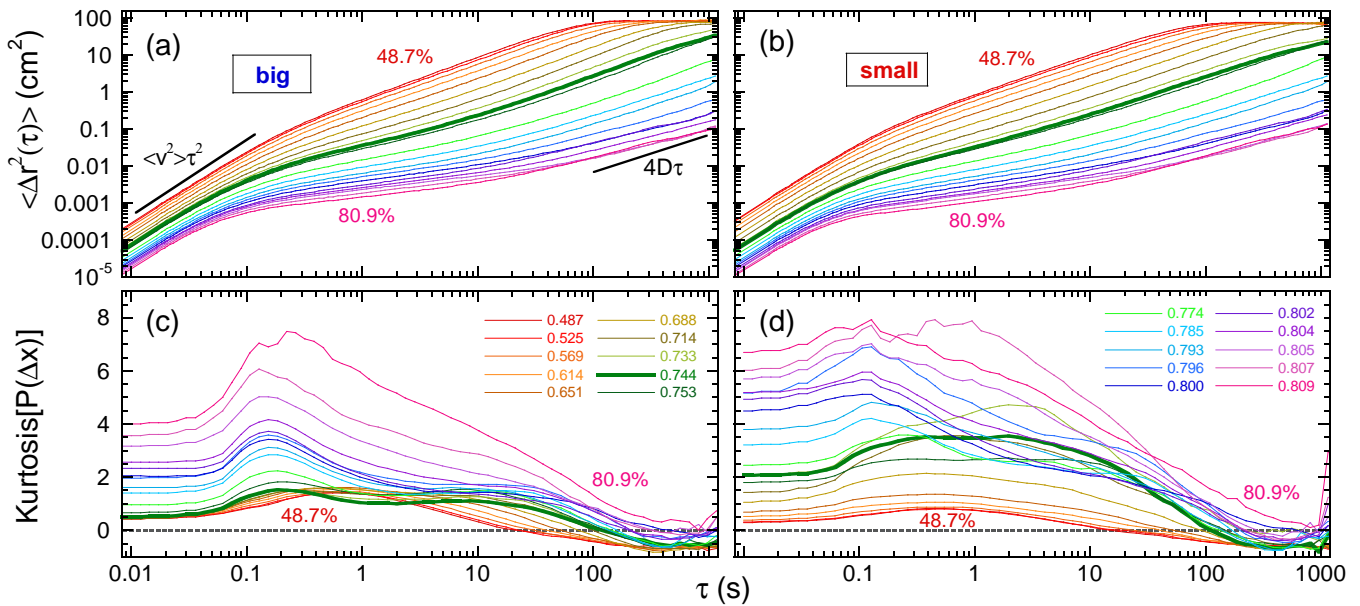}
\caption{(Color online) (a-b) Mean-squared displacement, and (c-d)
kurtosis of displacement probability distribution, all as a function
of delay time.  The left-hand plots are for the big beads and the
right-hand plots are for the small beads. Area fraction color codes
for the all plots are labeled in (c-d); the thick green curve is for
$\phi=74.4\%$.  Note that the mean-squared displacement saturates at
the square of the sample cell radius.  The squares of bead diameters
are $d_b^2=0.76~{\rm cm}^2$ and $d_s^2=0.40~{\rm cm}^2$.  The square
of the position resolution is $(0.0016~{\rm
cm})^2=3\times10^{-6}~{\rm cm}^2$.}\label{msd}
\end{figure*}

\begin{figure}
\includegraphics[width=3.30in]{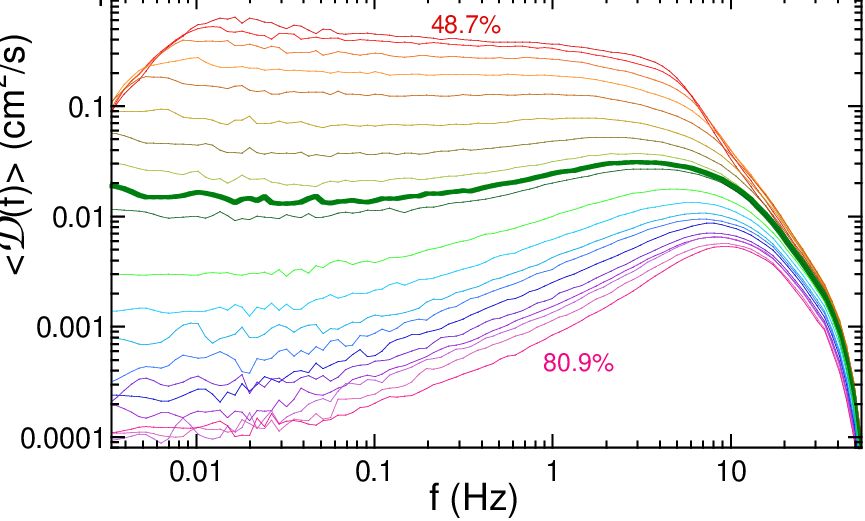}
\caption{(Color online) Density of vibrational states of frequency
$f$, for various packing fractions; the thick green curve is for
$\phi=74.4\%$. }\label{Df}
\end{figure}

For the remainder of the paper we focus on bead motion, and how it
changes in response to the structural changes found above an area
fraction of $74\%$. The primary quantity we measure and analyze is
the mean-squared displacement (MSD) that the beads experience over a
time interval $\tau$: $\langle \Delta r^2(\tau)\rangle$ vs $\tau$.
The MSD is readily measured directly from time- and ensemble
averages of the position vs time data; it can also be computed
efficiently from position autocorrelation data using Fourier
methods.  Results are shown in Figs.~\ref{msd}a-b for the big and
small beads, separately. At short times, the bead motion is
ballistic and characterized by a mean-squared speed according to
$\langle \Delta r^2(\tau)\rangle = \langle v^2\rangle \tau^2$. Our
frame rate is 120~Hz, corresponding to a shortest delay time of
$\tau=0.0083$~s, which is fast enough that we are able to observe
ballistic motion over about one decade in time for all area
fractions. At long times, the bead motion is diffusive and
characterized by a diffusion coefficient according to $\langle
\Delta r^2(\tau)\rangle = 4D\tau$.  Our run durations are
20~minutes, corresponding to a longest delay time of $\tau=1200$~s,
which are long enough for the beads to explore the entire system
several times at the lowest area fractions.  The crossover from
ballistic to diffusive regimes becomes progressively slower as the
area fraction increases.  Our run durations are long enough to fully
capture the diffusive regime at all but the highest area fractions.
Thus our full position vs time dataset, for all beads and area
fractions, should suffice for a complete and systematic study of
changes in dynamics as jamming is approached.

The MSD has long been used to characterize complex dynamics. In
simple systems there is a single characteristic time scale given by
the crossover from ballistic to diffusive regimes. In supercooled or
glassy systems, the crossover is much more gradual and there are two
characteristic time scales.  The shortest, called the `$\beta$'
relaxation time, is given by the end of the ballistic regime.  The
longest, called the `$\alpha$' relaxation time, is given by the
beginning of the diffusive regime. At greater degrees of
supercooling in glass-forming liquids, and at greater packing
fractions in colloidal suspensions, the $\alpha$ relaxation time
increases and a corresponding plateau develops in the MSD.  As seen
in our MSD data of Figs.~\ref{msd}a-b, this same familiar sequence
of changes occurs in our system as well. The greater the delay in
the onset of diffusive motion, the more `supercooled' is our system
and the closer it is to being jammed -- where each bead has a fixed
set of neighbors that never changes.

Another similarity between the dynamics in our system and thermal
systems can be seen by examining the kurtosis of the displacement
distribution. For a given delay time, there is a distribution
$P(\Delta x)$ of displacements. By symmetry the average displacement
and other odd moments must all vanish.  The second moment is the
most important; it is the MSD already discussed. If the distribution
is Gaussian, a.k.a.\ normal, then all other even moments can be
deduced from the value of the MSD. For example the `kurtosis' is the
fourth moment scaled by the square of the MSD and with the Gaussian
prediction subtracted; by construction it equals zero for a Gaussian
distribution, and otherwise is a dimensionless measure of deviation
from `normality'.  The kurtosis of the displacement distribution has
been used in computer simulation of liquids, both
simple~\cite{RahmanPR64} and supercooled~\cite{ThirumalaiPRE93,
HarrowellJCP96, GlotzerPRL97}, as well as in
scattering~\cite{vanmegen88} and imaging experiments of dense
colloidal suspensions~\cite{SillescuL98, MarcusPRE99, KegelSCI00,
WeeksSCI00}.  These works consider a quantity $\alpha_2$ equal to
$1/3$ of the kurtosis, and find that the displacement distribution
is Gaussian in ballistic and diffusive regimes, but becomes
non-Gaussian with a peak in the kurtosis at intermediate times. We
computed the kurtosis of the displacement distribution for our
system, with results displayed vs delay time in Figs.~\ref{msd}c-d
for many area fractions.  As in thermal systems, our kurtosis
results display a peak at intermediate times and becomes
progressively more Gaussian at late times. Furthermore, the peak
increases dramatically once the area fraction rises above about
74\%, particularly for the large beads.

The kurtosis data in Figs.~\ref{msd}c-d do not vanish at short
times, by contrast with thermal systems. Instead, we find that the
kurtosis decreases to the left of the peak and saturates at a
nonzero constant upon entering the ballistic regime. At these short
times, the displacement distribution has the same shape as the
velocity distribution since $\Delta x=v_x\tau$. Indeed our velocity
distributions are non-Gaussian with the same kurtosis as the
short-time displacement distributions.  This reflects the
non-thermal, far-from-equilibrium, driven nature of our
air-fluidized system.  Another difference from thermal behavior, as
we'll show in the next subsection, is that the two bead species have
different average kinetic energies -- which is forbidden by
equipartition for a thermal systems.

While the second and fourth moments of the displacement
distributions capture many aspects of bead motion, another dynamical
quantity has been considered recently~\cite{OHernPRE03, LeoPRL05,
WyartEL05, WyartPRE05}: the density $\langle {\cal D}(f)\rangle$ of
vibrational states of frequency $f$.  At high frequencies, the
behavior of $\langle {\cal D}(f)\rangle$ reflects the short time
ballistic nature of bead motion.  At low frequencies, the behavior
of $\langle {\cal D}(f)\rangle$ reflects on slow collective
relaxations.  If the system is unjammed, there will be
zero-frequency translational relaxation modes with relative
abundance set by the value of $\langle {\cal D}(0)\rangle$. If the
system is fully jammed, by contrast, there can be no zero-frequency
modes and hence $\langle {\cal D}(f)\rangle$ must vanishes for
decreasing $f$.  Thus the form and the limit of $\langle {\cal
D}(f)\rangle$ at low frequencies give a sensitive dynamical
signature of the approach to jamming. This shall be our focus, while
by contrast in Refs.~\cite{OHernPRE03, LeoPRL05, WyartEL05,
WyartPRE05} the focus was on the behavior of low frequency modes
above close packing on approach to {\it un}jamming.

The density of states may be computed from the velocity time traces,
$v_i(t)$, for all beads $i$~\cite{Dove93}.  The mass-weighted
ensemble average of time-averaged velocity autocorrelations,
$w(\tau)=\sum m_i\langle {\bf v}_i(t) \cdot {\bf v}_i(t+\tau)\rangle
/ \sum m_i$, is the key intermediate quantity; its Fourier transform
is $w(f)$ and has units of cm$^2$/s. The final step is to compute
the modulus,
\begin{equation}\label{densityofstates}
    \langle {\cal D}(f)\rangle=\sqrt{w(f)w^*(f)}.
\end{equation}
The angle-brackets in this notation are a reminder that this is an
ensemble average of single-grain quantities.  By construction, the
integral of Eq.~(\ref{densityofstates}) over all frequencies is
equal to the mass-weighted mean-squared speed of the beads. While
Eq.~(\ref{densityofstates}) appears to be a purely mathematical
manipulation, the identification of the right-hand side with the
density of states requires that modes be populated according to the
value of $kT$; hence, for non-thermal systems like ours, the result
is an effective density of states that only approximates the true
density of states.  Whatever the accuracy of this identification,
both the expression for $\langle {\cal D}(f)\rangle$ in
Eq.~(\ref{densityofstates}) and the mean-squared displacement may be
computed from velocity autocorrelations, and thus do not embody
different physics; rather, they give complementary ways of looking
at the same phenomena and serve to emphasize different features.

The effective density of vibrational states for our system of
air-fluidized beads is shown in Fig.~\ref{Df}, with separate curves
for different area fractions.  At low $\phi$ there are many low
frequency modes and $\langle {\cal D}(f)\rangle$ is nearly constant
until dropping off at high frequencies.  At higher $\phi$, the
number of low frequency modes gradually decreases; $\langle {\cal
D}(f)\rangle$ is still constant at low $f$, but it increases to a
peak before dropping off at high $f$. Even at the highest area
fractions, $\langle {\cal D}(f)\rangle$ is nonzero at the smallest
frequencies observed, as given by the reciprocal of the run
duration.

\subsection{Short and long-time dynamics}

\begin{figure}
\includegraphics[width=3.30in]{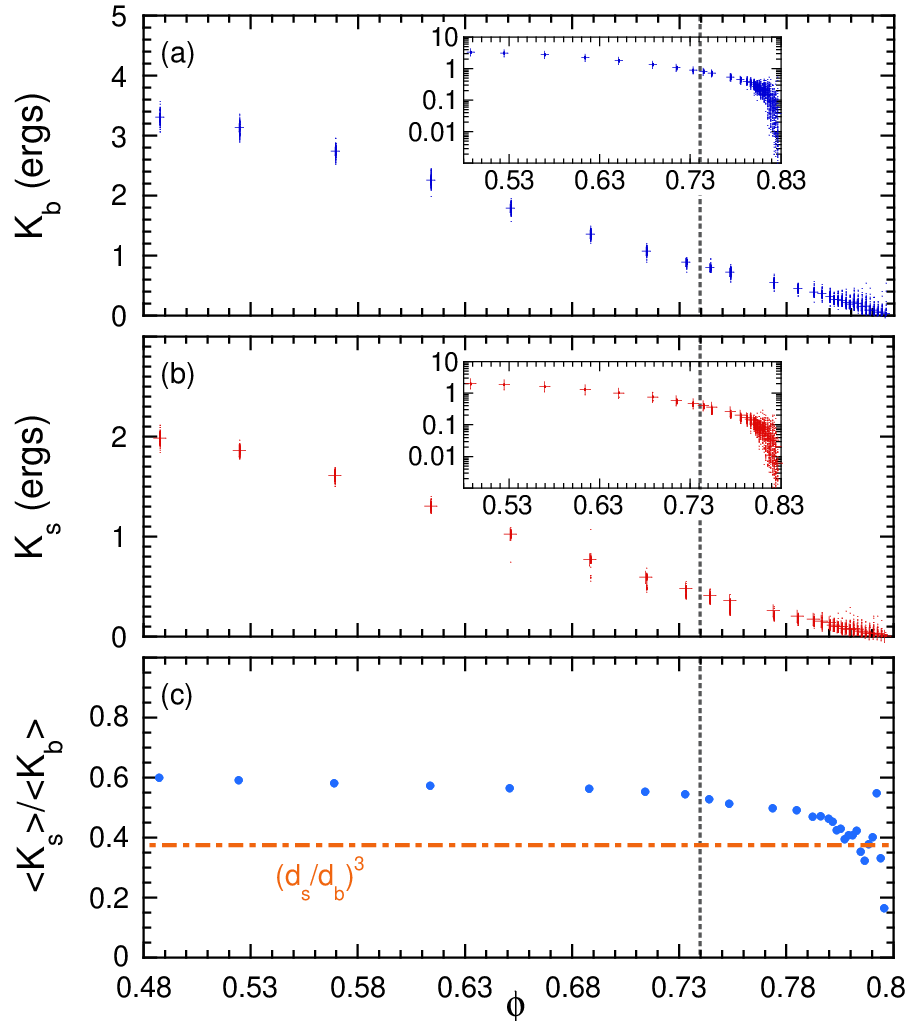}
\caption{(Color online) (a-b) Kinetic energy vs area fraction for
big and small beads, respectively.  The time-average for each bead
is shown by a small dot; the ensemble average over all beads is
shown by $+$. (c) The kinetic energy ratio of small-to-big beads.
The line $(d_{s}/d_{b})^{3}$ shows where beads move with the same
mean-squared velocity.  Insets show the same quantities on a
logarithmic scale.}\label{KKr}
\end{figure}

\begin{figure}
\includegraphics[width=3.30in]{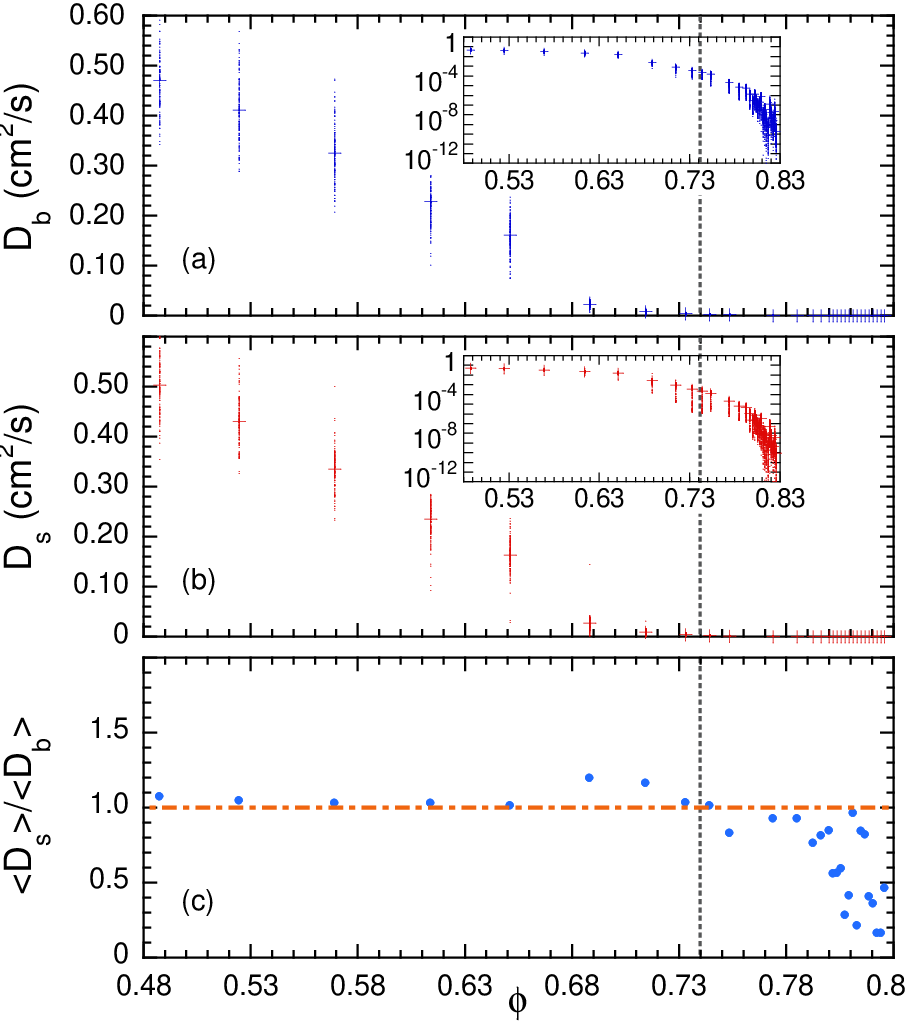}
\caption{(Color online) (a-b) Diffusion coefficient vs area fraction
for big and small beads, respectively.  The time-average for each
bead is shown by a small dot; the ensemble average over all beads is
shown by $+$.  (c) The diffusion coefficient ratio of small-to-big
beads.  Insets show the same quantities on a logarithmic
scale.}\label{D}
\end{figure}

The quantities that specify the ballistic and diffusive motion at
short and long times, respectively, are the mean-squared speeds and
the diffusion coefficients.  These may be deduced from the
mean-squared displacement data, separately for each bead. We now
examine trends in these dynamical quantities as a function of area
fraction.

We begin with the mean-squared speed.  To highlight the contrast
with a thermal system, we convert it to a mean kinetic energy and
plot the results in Figs.~\ref{KKr}a-b for the big and small beads,
respectively.  We show a small point for each individual bead, as
well as a larger symbol for the ensemble average of these values.
The average kinetic energies appear to decrease nearly linearly
towards zero as the area fraction is raised towards random close
packing. The scatter in the points is roughly constant, independent
of $\phi$, and reflects the statistical uncertainty in our velocity
measurements.  There is no evidence of nonergodicity or
inhomogeneity in energy injection by our air-fluidization apparatus;
namely, all the beads of a given species have the same average
kinetic energy to within measurement uncertainty. Beyond about
$\phi=0.81$ the uncertainty becomes comparable to the mean, as set
by our speed resolution, and we can no longer readily discern the
average kinetic energies. This may be more evident in the insets,
which show kinetic energies on a logarithmic scale.

Note that the big and small beads have different average kinetic
energies.  Equipartition is thus violated for our athermal, driven
system.  This is highlighted in Fig.~\ref{KKr}c, which shows the
kinetic energy ratio of small to large beads.  This ratio is roughly
constant at about 0.6 for the lowest area fractions.  It decreases
gradually for increasing $\phi$, at first gradually and then more
rapidly beyond about $\phi=0.78$.  There is no obvious feature at
$\phi=0.74$, where structural quantities changed noticeably.
Interestingly, however, the kinetic energy ratio appears to head
toward the cube of the bead diameter ratio on approach to random
close packing. This means that the mean-squared speeds are
approaching the same value, perhaps indicating that only extremely
collective motion is possible very close to jamming.  In order for
one bead to move, the neighboring bead in the path of motion must
move with the same speed.  This seems a natural geometrical
consequence of nearly close packing, but would be a violation of
equipartition in a thermal system.

Turning now to late-time behavior, we display diffusion coefficients
vs area fraction in Figs.~\ref{D}a-b, for big and small beads
respectively.  As in Fig.~\ref{KKr}, a small dot is shown for each
individual bead and a larger symbol is shown for the ensemble
average of these results.  The average diffusion coefficients appear
to decrease linearly with increasing area fraction, starting at the
lowest $\phi$.  Here there is considerable scatter in the data,
reflecting a level of uncertainty set by run duration. Linear fits
of diffusion coefficient vs area fraction in this regime extrapolate
to zero at the special value $\phi=0.74$. However, just before
reaching this area fraction, the data fall below the fit. As shown
on a logarithmic scale in the insets, the diffusion coefficients are
nonzero and continue to decrease with increasing $\phi$.  Beyond
about $\phi=0.81$ the system becomes non-ergodic over the duration
of our measurements. Namely, some beads remain stuck in the same
nearest neighbor configurations while others have broken out.
Indeed, at these high area fractions the MSDs in Fig.~\ref{msd} grow
sublinearly at the latest observed times.  To obtain reliable
diffusion coefficient data in this regime would require vastly
longer run durations.

\subsection{Timescales}

\begin{figure}
\includegraphics[width=3.30in]{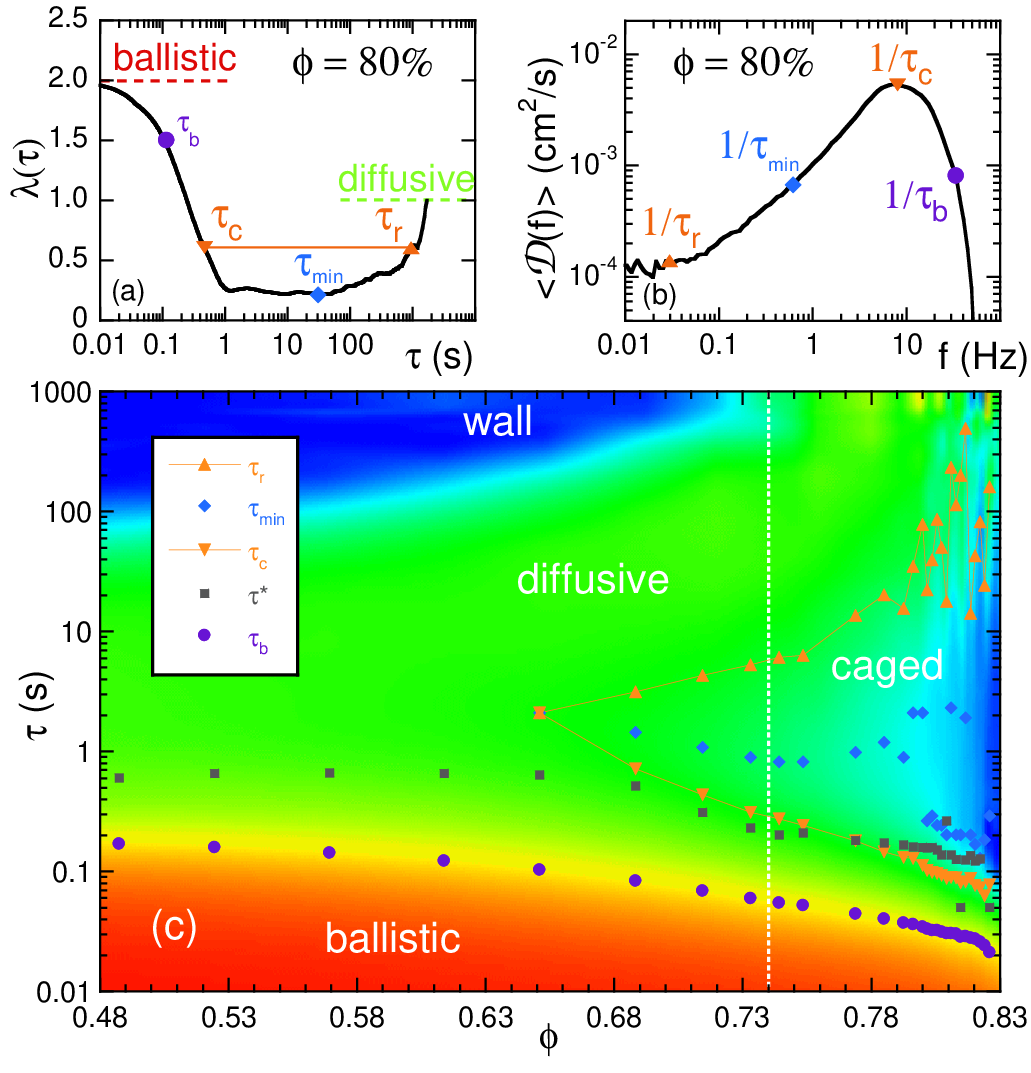}
\caption{(Color online) (a) Logarithmic derivative $\lambda$ of the
mean-squared displacement vs delay time, for big beads at area
fraction $\phi=80\%$. Four special time scales can be defined from
such data, as depicted: $\tau_b$ where $\lambda=1.5$, $\tau_{\rm
min}$ where $\lambda$ is minimum, and $\tau_c$ and $\tau_r$ where
$\lambda$ is halfway between 1 and its minimum.  (b) Density of
state, for big beads at $\phi=80\%$, marked with the four special
timescales. (c) Contour plot of the logarithmic derivative for the
big beads, where color indicates the value of $\lambda$, as a
function of both area fraction and delay time. Red is slope two and
blue is slope zero; caged dynamics are when the mean-squared
displacement has a slope less than one, which is aqua-blue. The four
special times defined by the behavior of $\lambda$, as well as a
fifth time $\tau^*$ at which the displacement kurtosis is maximal,
are superposed on the contour plot; symbol key is given in the
legend.}\label{Inflects}
\end{figure}

Lastly we focus on characterizing several special times that serve
to demarcate the short-time ballistic and the long-time diffusive
regimes seen in the displacement statistics data.  The time
scale-dependent nature of the dynamics, which is obvious in the MSD
or the density of states, is also obvious in real-time observations
or in AVI movie data.  For high area fractions, the bead
configuration appears immutable at first; the beads collide
repeatedly with an apparently-fixed set of neighbors that cage them
in.  Only with patience, after hundreds or thousands of collisions,
can the beads be observed to break out of their cages, change
neighbors, and begin to diffuse throughout the system.

To measure unambiguously the characteristic timescales, we consider
the slope of the mean-squared displacement as seen on a log-log
plot:
\begin{equation}\label{dogmsd}
    \lambda(\tau) = {\partial \ln[\langle\Delta r^2(\tau)\rangle]
                      \over  \partial \ln\tau}.
\end{equation}
An example for one of the higher area fractions is shown in
Fig.~\ref{Inflects}a.  At the shortest delay times, when the motion
is perfectly ballistic, this slope is 2; at the longest delay times,
when the motion is perfectly diffusive, this slope is 1.  For high
area fractions, such as shown, there is a subdiffusive regime with
$\lambda<1$ at intermediate times; this is when a typical bead
appears by eye to be stuck in a cage, rattling against a fixed set
of neighbors. For low area fractions, not shown, there is no such
`caging' and $\lambda$ decreases monotonically from 2 to 1.

Several natural time scales can now be defined with use of
$\lambda(\tau)$ vs $\tau$ data.  The shortest is the delay time
$\tau_b$ at which the logarithmic slope falls to
$\lambda(\tau_b)=1.5$. This demarcates the ballistic regime, below
which the bead velocity is essentially constant.  At high area
fractions, $\tau_b$ is a typical mean-free time between successive
collisions; at low area fractions, $\tau_b$ it is also the time for
crossover to diffusive motion.  The other time scales that we define
all refer to the subdiffusive, caging dynamics at high area
fractions.  The most obvious is the delay time $\tau_{\rm min}$ at
which $\lambda(\tau_{\rm min})$ is minimum; this corresponds to an
inflection in the MSD on a log-log plot.  Below $\tau_{\rm min}$
most beads remain within a fixed cage of neighbors.  The last two
special times specify the interval when the motion is subdiffusive,
with a logarithmic slope falling in the range below 1 and above its
minimum.  The smaller is the delay time $\tau_c$ at which the
logarithmic slope decreases half-way from 1 down towards its
minimum: $\lambda(\tau_c) = [1 + \lambda(\tau_{\rm min})]/2$. This
is the time at which the beads have explored enough of their
immediate environment to `realize' they are trapped at least
temporarily within a fixed cage of neighbors; it is longer, and
distinct from, the mean-free collision time.  The longest special
time scale is the delay time $\tau_r$ at which the logarithmic slope
increases half-way from its minimum up towards 1: $\lambda(\tau_r) =
[1 + \lambda(\tau_{\rm min})]/2$. This is the time beyond which the
beads rearrange and break out of their cages; it demarcates the
onset of fully diffusive motion.

Before continuing, we note that these four special times also
correspond to features in the density of states.  As shown in
Fig.~\ref{Inflects}b, the density of states drops to zero
precipitously for frequencies above $1/\tau_b$, the reciprocal of
the ballistic- or collision mean-free time.  The density of states
reaches a peak at $1/\tau_c$, corresponding to the time that beads
`realize' they are stuck at least temporarily within a cage.  For
lower frequencies, the logarithmic derivative of $\langle {\cal
D}(f)\rangle$ vs $f$ has an inflection at roughly $1/\tau_{\rm
min}$. And at the lowest frequencies, $\langle {\cal D}(f)\rangle$
approaches a constant value below about $1/\tau_r$.  Thus we could
have analyzed all data in terms of $\langle {\cal D}(f)\rangle$;
however we prefer to work with the mean-squared displacement since
it does not involve numerical differentiation.

Results for the special timescales are collected in
Fig.~\ref{Inflects} as a function of area fraction.  Actual data
points are superposed on top of a contour plot of the logarithmic
derivative, where color [on-line] fans through the rainbow according
to the value of $\lambda$: red for ballistic, green for diffusive,
and blue for subdiffusive. With increasing $\phi$, the ballistic
time scale decreases steadily by a factor of nearly ten as close as
we can approach random close packing.  Note that with our definition
of $\tau_b$, these data points lie near the center of the yellow
band demarcating the end of the red ballistic regime. Below
$\phi=0.65$ the motion is never subdiffusive, and $\tau_b$ is the
only important time scale.  Right at $\phi=0.65$ the motion is only
barely subdiffusive, for a brief moment, so that all three
associated times scales nearly coincide.  Above $\phi=0.74$ the
caging is sufficiently strong that the time scale for rearrangement
$\tau_r$ is ten times longer than the time scale $\tau_c$ for
`realization' that there is caging.  At progressively higher area
fractions this separation in time scales grows ever stronger.  On
approach to jamming at random close packing, the cage realization
time decreases towards a nonzero constant, while the cage break-up
or rearrangement time $\tau_r$ increases rapidly towards our run
duration and appears to be diverging.

Note that the two time scales $\tau_c$ and $\tau_r$ capture the
subdiffusive caging dynamics better than just the delay time
$\tau_{\rm min}$ at which the logarithmic derivative is minimum and
the motion is maximally subdiffusive.  One reason is that $\tau_{\rm
min}$ is difficult to locate for extremely subdiffusive motion,
where there is a wide plateau in the MSD and hence where $\lambda$
is nearly zero over a wide range of delay times.  This difficulty
can be seen in both Fig.~\ref{Inflects}a, where $\lambda$ remains
close to its minimum for over two decades in delay time, as well as
in Fig.~\ref{Inflects}c above about $\phi=0.8$, where $\tau_{\rm
min}$ data jump randomly between about 2~s and about $3\tau_c$.  The
other reason is that $\tau_{\rm min}$ does not appear to be either a
linear or geometric average of the cage `realization' and breakup
times.  Rather, the shape of $\lambda(\tau)$ vs $\ln(\tau)$ is
asymmetric, with a minimum closer to the cage `realization' time.
Thus it is useful to know the two times $\tau_c$ and $\tau_r$ that
together specify the subdiffusive dip of $\lambda(\tau)$ below one,
just as it is useful to know the full-width half-max of a spectrum
of unspecified shape.

Note too that $\tau_c$ and $\tau_r$ are distinct from the time
$\tau^*$ at which the kurtosis is at maximum. Results for $\tau^*$
are extracted from our kurtosis data, and are displayed as open
squares along with other characteristic times in
Fig.~\ref{Inflects}.  At low area fractions, even when there is no
subdiffusive regime, the kurtosis exhibits a maximum at delay time
$\tau^*$ that is several times the ballistic / collision time
$\tau_b$.  With increasing area fraction, $\tau^*$ decreases.  Once
a subdiffusive regime appears, the value of $\tau^*$ is close to the
time $\tau_c$ at which grains `realize' they are stuck in a cage.
Data in Fig.~\ref{Inflects} for both $\tau_c$ and $\tau^*$ decrease
with increasing area fraction, $\phi$, on approach to jamming, while
the time $\tau_r$ signalling the end of the subdiffusive regime
appears to grow without bound. The decrease of $\tau^*$ with $\phi$
agrees with previous observations on colloids~\cite{MarcusPRE99,
KegelSCI00, WeeksSCI00}, but contrasts with statements that $\tau^*$
corresponds to the cage-breakout $\alpha$-relaxation time beyond
which the motion is diffusive.  To emphasize, we find that the tail
of the displacement distribution is largest relative to a Gaussian,
and hence that the kurstosis is maximal, at the {\it beginning} of
the subdiffusive regime. At that time most beads have been turned
back by collision, but a few prolific beads move at ballistic speed
roughly one diameter -- which is long compared to the rms
displacement.  This observation is not an artifact of limited
spatial or temporal resolution or of limited packing fraction range,
which are all notably better than in previous experiments. It is
also not an artifact of our analysis method; indeed, if we filter
too strongly then the kurtosis peak shifts to later times.

\subsection{Jamming phase diagram}

\begin{figure}
\includegraphics[width=3.30in]{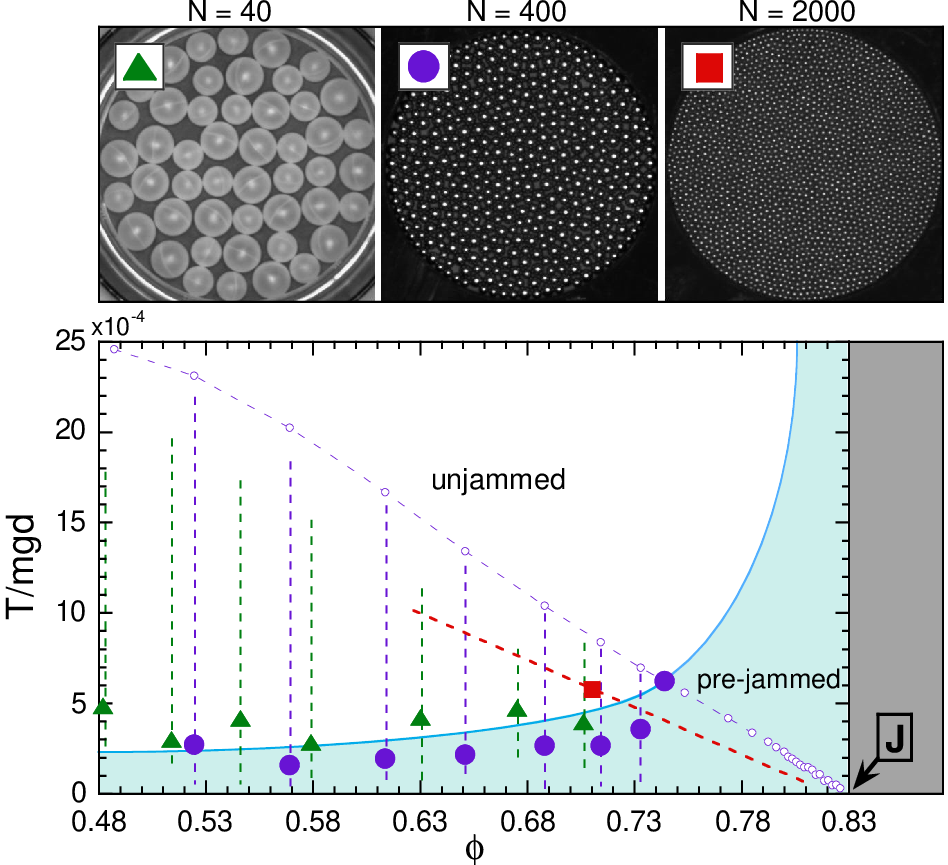}
\caption{(Color online) The zero-stress plane of the jamming phase
diagram showing our trajectories in phase space.  The primary
trajectory, which corresponds to sequences shown in all previous
plots, is given by the diagonal dashed purple curve and open circles
that intersects the ``pre-jammed'' boundary at $74\%$ and that ends
at Point J, $\{\phi\approx0.83,T=0\}$.  The grayed region is
forbidden for impenetrable beads. The solid circles, triangles, and
square denote measurements of the phase boundary for large hollow
polypropylene beads, solid steel beads, and tiny solid steel beads,
as shown in the images. Each boundary point corresponds to the
trajectory indicated by the dashed going through it.}\label{PointJ}
\end{figure}

Before closing we now summarize our structure and dynamics
observations and place them in the context of the jamming phase
diagram.  By adding more beads to our system at fixed air speed, we
control the approach to jamming two ways. First, most obviously, the
area fraction increases towards random close packing at
$\phi_c=0.83$. Second, the average kinetic energy of the beads
decreases -- roughly linearly with $\phi$ according to the kinetic
energy data shown in Fig.~\ref{KKr}. Both aspects are captured by
the trajectory on a jamming phase diagram plot of kinetic energy vs
area fraction. To get a single dimensionless measure of kinetic
energy, we divide the mass-weighted average kinetic energy of the
beads by a characteristic gravitational energy given by average ball
weight times diameter; the result is denoted as a scaled effective
temperature $T/(mgd)$. This scaling reflects the natural energy for
air-mediated interactions in a gas-fluidized system.  The particular
$T/(mgd)$ vs $\phi$ trajectory followed by our experiments is shown
by open symbols in Fig.~\ref{PointJ}.  It is a diagonal line that
terminates at $\{T=0,\ \phi=\phi_c\}$, which is the special point in
the jamming phase diagram known as `Point-$J$' \cite{OHernPRL02,
OHernPRE03}. On approach to Point-$J$ our system remains unjammed,
but develops tell-tale features in the Voronoi tessellations and in
the mean-squared displacement indicating that jamming is near. In
particular the structural changes saturate, and the ratio
$\tau_r/\tau_c$ of cage breakup to `realization' times exceeds ten,
for points along the trajectory closer to Point-$J$ than
$\{T/(mgd)=0.007, \phi=0.74\}$; we then say the system is
`pre-jammed'.  In this pre-jammed region, we did not observe any
signs of aging; the structure and dynamics change from that of a
simple liquid but do not appear to evolve with time.

An extended region of `pre-jammed' behavior must exist near
Point-$J$, and can be mapped out by changing the experimental
conditions.  The simplest variation is to examine vertical
trajectories, at fixed $\phi$ for the same system of steel beads,
where the effective temperature is changed by the speed of the
upflowing air.  We did this for several different area fractions,
locating the special effective temperatures below which the system
becomes pre-jammed. Furthermore, to better test the universality of
our phase diagram and the choice of energy scaling, we examined two
other 1:1 bidisperse mixtures of spheres. This includes several
constant-$\phi$ trajectories for large hollow polypropylene spheres
with diameters of 1-1/8 and 1-3/8 inches, and one constant air speed
trajectory for very small steel spheres with diameters of 5/32 and
1/8 inches. The resulting trajectories and pre-jamming boundary
points are shown in Fig.~\ref{PointJ} as dashed lines and symbols,
respectively.  In spite of the vast differences in bead systems, a
remarkably consistent region of pre-jammed behavior appears in the
temperature vs area fraction jamming phase diagram.

%=========================================================================================
\section{Conclusion}

The quasi-2D system of air-fluidized beads studied in this paper is
fundamentally different from an equilibrium system.  Here all
microscopic motion arises from external driving, and has nothing to
do with thermodynamics and ambient temperature.  Rather there is a
constant input of energy from the fluidizing air, and this excites
all motion. As a result of thus being far from equilibrium, the
velocity distributions are not Gaussian.  Furthermore, the average
kinetic energy of the two species is not equal because of how they
interact with the upflow of air.

In spite of these differences, our system exhibits hallmark features
upon approach to jamming that are very similar to the behavior of
thermal systems.  In terms of structure, our system develops a split
peak in the circular factor distribution and a split second peak in
the pair distribution function.  In terms of dynamics, our system
develops a plateau in the mean-squared displacement at intermediate
times, between ballistic and diffusive regimes, where the beads are
essentially trapped in a cage of nearest neighbors.  And also like a
thermal system, the prominence of these features increases with both
increasing packing fraction and with decreasing particle energy, in
a way that can be summarized by a jamming phase diagram.

The significance of the above conclusions is to help reinforce the
universality of the jamming concept beyond just different types of
thermal systems, to a broader class of non-equilibrium systems as
well.  This suggests that the geometrical constraints of disordered
packing plays the major role.  Our system of air-fluidized beads may
now serve as a readily-measured model in which to study further
aspects of jamming that are not readily accessible in thermal
systems.  For example it should now be possible to characterize
spatial heterogeneities and dynamical correlations in our system,
expecting the results to shed light on all systems, thermal or not,
that are similarly close to being jammed.

\begin{acknowledgments}
We thank Andrea Liu and Dan Vernon for helpful discussions, and Mark
Shattuck for introducing us to the shape factor in
Ref.~\cite{MouckaPRL05}. Our work was supported by NSF through Grant
No.~DMR-0514705.
\end{acknowledgments}

% Create the reference section using BibTeX:
\bibliography{multirefs}

\end{document}